\documentclass[10pt, conference]{IEEEtran}
\usepackage{amsfonts}
\usepackage{amssymb}
\usepackage{eurosym}
\usepackage{cite}
\usepackage{graphicx}
\usepackage{epstopdf}
\usepackage{amsmath}
\usepackage{tikz,lipsum}
\usepackage{caption}
\usepackage[T1]{fontenc}
\usepackage{amsthm}
\usepackage{mathrsfs}
\usepackage{color}
\usepackage{stfloats}
\usepackage{xcolor, etoolbox}
\usepackage{subcaption}
\usepackage{authblk}
\usepackage{comment}
\usepackage{amsmath}
\usepackage{algorithm}
\usepackage{algorithmic}
\usepackage{balance}
\usepackage{hyperref}

\makeatletter
\def\ps@IEEEtitlepagestyle{%
  \def\@oddfoot{\mycopyrightnotice}%
}
\def\mycopyrightnotice{%
  \begin{minipage}{\textwidth}
  \centering \scriptsize
  \copyright 2024 IEEE. Personal use of this material is permitted. Permission from IEEE must be obtained for all other uses, in any current or future media, including reprinting/republishing this material for advertising or promotional purposes, creating new collective works, for resale or redistribution to servers or lists, or reuse of any copyrighted component of this work in other works.
  \end{minipage}
}
\makeatother

\IEEEoverridecommandlockouts

\def\BibTeX{{\rm B\kern-.05em{\sc i\kern-.025em b}\kern-.08em
    T\kern-.1667em\lower.7ex\hbox{E}\kern-.125emX}}

\begin{document}

\title{Elevating the Future of Mobility: UAV-enabled Intelligent Transportation Systems\\
}

\author{Abdul Saboor\textsuperscript{1},
        Evgenii Vinogradov\textsuperscript{1, 2},
        Zhuangzhuang Cui\textsuperscript{1},
        Sander Coene\textsuperscript{3},
        Wout Joseph\textsuperscript{3},
        Sofie Pollin\textsuperscript{1}
\\\textsuperscript{1}WaveCoRE of the Department of Electrical Engineering (ESAT), KU Leuven, Leuven, Belgium
\\\textsuperscript{2}Autonomous Robotics Research Center, Technology Innovation Institute, Abu Dhabi, UAE
\\\textsuperscript{3}Department of Information Technology, IMEC-Ghent University, 9052 Ghent, Belgium
\\Email:\{abdul.saboor, zhuangzhuang.cui, sofie.pollin\}@kuleuven.be, evgenii.vinogradov@tii.ae, \\\{sander.coene, wout.Joseph\}@ugent.be

}


\maketitle

\begin{abstract}
Intelligent Transportation Systems (ITS) increasingly rely on connectivity for efficient traffic management and enhanced user experience. The existing ITS solutions operate mainly within a 2D domain, thus missing the potential benefits of aerial platforms. This paper envisions 3D ITS by integrating aerial platforms, such as Unmanned Aerial Vehicles (UAVs), to simultaneously improve network coverage and support multi-modal transportation, including Advanced Air Mobility (AAM). Using stochastic models, we investigate how UAV-based Aerial Base Stations (ABSs) can address the limitations of traditional Terrestrial Base Stations (TBSs) by offering superior coverage, particularly in urban environments. Our results demonstrate that ABSs have 106.67\% more coverage area than TBS, higher Signal-to-Noise Ratio (SNR) distribution, and are suitable for high-throughput ITS applications.
\end{abstract}

\begin{IEEEkeywords}
Intelligent Transportation Systems (ITS), Aerial Base Stations (ABS), Unmanned Aerial Vehicles (UAVs), Advanced Air Mobility (AAM)
\end{IEEEkeywords}

\section{Introduction}
A significant rise is observed this century in urbanization, with more people migrating to urban areas. This influx of new residents increases the size of cities around the globe, resulting in higher transportation demands, traffic congestion, and pollution \cite{Rashed23}. In response to these challenges, the mobility sector is undergoing transformative changes. Intelligent Transportation System (ITS) is one of the revolutions to address such challenges that includes driverless cars, cooperation between vehicles (e.g., supported by 5G networks), electric cars to reduce pollution, and shared individual transport (e.g., cars, bikes, or scooters). While the ITS represents a significant step forward, an even more revolutionary shift is underway.

Advanced Air Mobility (AAM) envisions a safe and efficient transportation system using highly automated aircraft to operate and transport passengers or cargo at lower altitudes within urban and suburban areas \cite{10472699}. In May 2021, Morgan  Stanley released a prediction \cite{MS} stating that by 2050, the total market of AAM (delivery, air taxi, patrolling drones, to name a few) will reach up to \$19 tn (10-11\% of projected global Gross Domestic Product (GDP)). Furthermore, drone delivery will become part of urban life within the next fifteen years, and the commercial introduction of urban passenger aircraft will take around 20-30 years.

The same 3-dimensional (3D) shift is observed in the telecommunication sector \cite{UAV_6G}. One of the 6G enablers is Non-Terrestrial Networks (NTNs), including Unmanned Aerial Vehicles (UAVs), satellites, High Altitude Platforms (HAPs), and ground infrastructure \cite{Oliveri2024,10299705}. AAM and NTN reinforce and help each other, creating new use cases and attracting much attention from academia and industry. In contrast, ITS lacks the support of aerial infrastructure. Its reliance on terrestrial infrastructure and road-bound vehicles poses several challenges. For example, Terrestrial Base Stations (TBS) suffer from coverage gaps and higher path loss due to lower Line-of-Sight (LoS) availability, particularly in dense urban areas with various obstacles \cite{10333093, AWPL}. Furthermore, the fixed nature of TBS limits the flexibility and adaptability of ITS systems. Similarly, the reliance on road-bound vehicles can hinder the development of innovative mobility solutions that extend beyond traditional road-based transportation.  

To address the challenges of existing (2D) ITS, this paper proposes integrating AAM and NTN to propose a novel paradigm called 3D ITS. With a major focus on UAVs in NTN, the proposed framework aims to revolutionize urban mobility and connectivity by combining the strengths of ground-based and aerial transportation systems. The overall contributions of this paper are:

\begin{itemize}
    \item We envision a 3D ITS that integrates 2D ITS, AAM, and NTN and outlines its potential use cases.
    \item We use Poisson processes to model vehicles and base stations for performance evaluation of Aerial Base Stations (ABS) and TBS in urban environments. Our findings highlight the significant connectivity advantages that an ABS can provide to the ITS.   
\end{itemize}

The rest of the paper is organized as follows. Section II overviews traditional 2D ITS. Section III identifies the UAV-enabled 3D ITS and its use cases. Section IV compares the performance of ABS and TBS in urban environments. Finally, Section V concludes our work. 

\begin{figure*}[!t]
\begin{center}
  \includegraphics[width=1\linewidth]{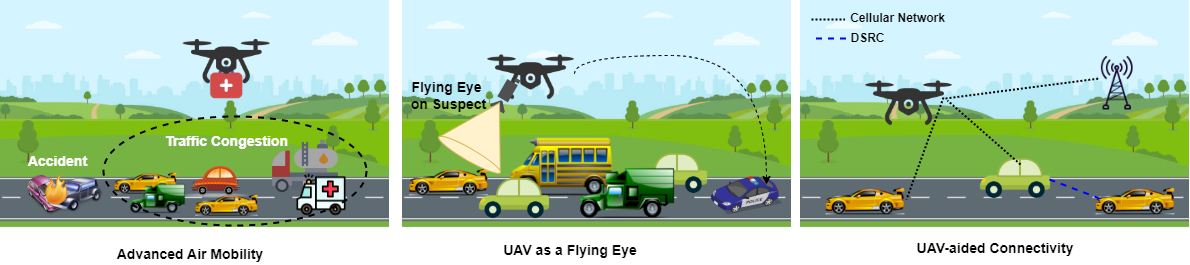}
  \caption{UAV use cases in ITS: a) a UAV is providing early medication by skipping traffic congestion, b) a UAV is chasing a suspect and sending information to the police, c) a UAV is acting as ABS to provide connectivity in road vicinity. }
  \label{usecases}
  \end{center}
\end{figure*}

\section{Overview of 2D Intelligent Transportation Systems} 
ITS aims to reduce friction for participants in a transportation scenario. At the end of the last century, ITS saw independent development in the USA, China, Japan, and countries inside Europe, each spear-pointing different priorities. Early usage scenarios include navigation, which combines the Global Positioning System (GPS) (or any other Global Navigation Satellite System (GNSS)) with a digital roadmap to provide real-time instructions to get to a set destination. These ITS also contain automated toll systems using Radio Frequency Identification (RFID) chips or license plate recognition systems to identify cars and reduce the time needed for the payment.

Modern ITS relies on interconnected sensors and cameras to monitor traffic, assess vehicle types, detect congestion, and track visibility and accidents in real-time. Similarly, the connectivity enables dynamic road signs to warn vehicles of adverse road or traffic conditions, imposing suitable speed limits. In public transportation, seamless connectivity ensures real-time tracking, providing users with up-to-date information on arrival times and vehicle occupancy via apps and digital signs. In short, ITS requires automating the infrastructure, supporting units, devices, and networks to enable different services such as congestion control, collision avoidance, traffic light control, and parking \cite{10375912}. 

ITS has significantly improved transportation through advanced communication and automation. However, its dependence on terrestrial infrastructure and road-bound vehicles limits coverage, transportation flexibility, and adaptability, especially in dense urban environments. Integrating AAM and NTN in ITS can overcome these challenges and revolutionize urban mobility using aerial platforms operating in 3D space. 

\section{Next Generation 3D Intelligent Transportation System}
Menouar et al. \cite{7876852} demonstrated that UAVs help ITS from the wireless communications point of view because of mobility and better/stable radio channels. This section describes several ways of using UAVs to assist ITS through their transportation, surveillance, and connectivity capabilities.  
\subsection{Sky is the Limit: Advanced Air Mobility}
The transportation paradigm shifts towards 3D mobility will drastically change the traffic patterns, demand, and required infrastructure. In the following, we provide examples of the potential applications.
\subsubsection{Cargo Drones}
\paragraph{Emergency} Small drones can carry necessary medications or blood to emergency locations by avoiding congestion over the roads, as shown in Fig \ref{usecases}. A study conducted by the American Heart Association shows that exposure to traffic increases the chances of a cardiac arrest \cite{emergency}. The traffic caused approximately 8\% of cardiac arrest cases. In this case, a UAV can quickly deliver an Automated External Defibrillator (AED) by avoiding congestion to help restart the heart rate.
\paragraph{Road Congestion-friendly eCommerce} With the rise in online purchasing behavior, traffic congestion caused by delivery vehicles has also increased. As a result, research is being done on reducing the traffic load by using drones as part of the delivery system. Both UAV-only and truck-plus-UAV have already been studied as improvements \cite{mulumba2024optimization}. 
\paragraph{Temporary ITS Infrastructure}
UAVs can act as temporary signs in a particular area. The acquired data can be analyzed using machine learning algorithms to examine and predict traffic behavior in the presence of various signs. Finally, the fixed infrastructure (signs, roads) will be installed in suitable locations. Another advantage of adding UAVs to ITS is the ability to rapidly provide supporting material to remote positions, including fuel, aid, or toolkits.
\subsubsection{People Transportation}
\paragraph{Air Taxis}
Commercial AAM operations will help to avoid transportation congestion on the ground. Companies around the world are currently providing these services. For example, BLADE Bounce is an on-demand helicopter transfer service between BLADE helipad locations and New York area airports—both commercial and private. Similarly, Airbus is operating its Voom on-demand helicopter service in Sao Paulo, Brazil. However, the new all-electric and hybrid electric vertical take-off and landing (eVTOL) promises to be more accessible and efficient. These vehicles are currently being designed by Wisk (Boeing), Eve (Embraer), Airbus, Joby, Volocopter, and other companies.
\paragraph{Emergency}
In addition to passenger flights, AAM also includes aircraft operations for tasks in metropolitan areas, such as public safety, medical evacuations, and rescue.

\subsection{Bird-eye View: UAV as a Flying Eye}
UAV flying capability, equipped with advanced sensor and camera technology, offers various ITS use cases to enhance transportation safety, efficiency, and management.
\paragraph{Emergency}
UAVs can quickly reach an accident location and analyze the situation's severity using visual information. Furthermore, UAVs store and analyze traffic information in their maps. Therefore, they can reroute the ambulance to the emergency spot using the optimal route. The concept of a drone-assisted ambulance service can be another interesting future direction, where a drone takes the lead in managing traffic and signals using cellular technology to minimize congestion for the ambulance. 

\paragraph{Law Enforcement}
Police are an essential part of the ITS. The key duties of police include protection, law enforcement, patrolling, and violation tracking. UAVs can help the police enforce rules with a better aerial view than standard police vehicles. Similarly, the UAV-based police unit tracks traffic violations such as speed and wrong direction. Because of better speed and visibility, a UAV is also beneficial in chasing scenarios. Lastly, a UAV with a thermal sensor can assist the search and rescue missions. 


\paragraph{UAV-aided Traffic Monitoring and Management}
Due to the bird-eye view and cameras working in different parts of the spectrum, UAVs enable safer road use by pedestrians, cyclists, and Powered Two-Wheeler (PTW) riders (motorcycles, mopeds, etc.), collectively referred to as Vulnerable Road Users (VRU). According to the World Health Organization, VRUs represent the majority of road traffic fatalities, with 28\%, 23\%, and 3\% of deaths for PTW, pedestrians, and cyclists, respectively \cite{i2}. The factors affecting VRU safety demonstrated high socio-economic dependency (e.g., most causalities occur in developed countries' urban/rural environments). However, the most dangerous maneuver is a road crossing, and the most dangerous time of the day is twilight or darkness. 

\paragraph{ITS Infrastructure Inspection and Public Service}
UAVs can play an important role in enhancing ITS by enabling efficient inspection of traffic signalization and roadside conditions. They can assess damage to traffic signs, poles, and retroreflective markings, providing detailed visual analysis of parameters like serial numbers, material, dimensions, and light color. Furthermore, UAVs can monitor roadside conditions, inspect power lines, bridges, and pavements, create 3D building scans, track construction progress, and improve city planning. 



\subsection{Help From the Skies: UAV-aided Connectivity}
ITS enables vehicles to interact with each other and the surrounding road infrastructure to improve road safety and optimize transport efficiency. Despite significant work in this area, vehicular communication in the urban environment is still severely affected by the harsh propagation environment (e.g., dynamic obstacles). UAVs can offer a more stable communication link due to the flight altitude, ensuring persistent LoS channel \cite{saboorPlos,vin_tut}.
\paragraph{Flying Roadside Unit (RSU)}
RSU is important in ITS; its distribution directly influences vehicular communication quality. However, intermittent connectivity, high routing overhead, inflexible communication infrastructure, and unscalable architecture are the key challenges that hinder the wide applications of vehicular networks. Overcoming these hurdles is possible using UAV-mounted RSUs due to their inherent features like more stable channels (LoS link over longer distances) and dynamic on-demand deployment.
\paragraph{Aerial Base Stations (ABSs)}
Recently, the trend of relying on cellular networks for ITS support is gaining much attention. For example, the 5G Automotive Association (5GAA) indicates 5G as the next key technology for ITS \cite{5gaa2019bmac}. While cellular networks provide reliable communication between vehicles and infrastructure, fixed terrestrial infrastructure still faces coverage gaps and limited flexibility, similar to challenges with fixed RSUs. Deploying UAV-based ABS has emerged as a promising solution to address these limitations. UAV-based ABS offers dynamic coverage and adaptability, enhancing connectivity and performance for ITS applications. An overview of relevant ABS deployment techniques is provided in \cite{vin_tut}. However, the existing literature still needs a comprehensive analysis of ABS usage for vehicular communications in urban environments, which is addressed in the next section. 

\section{Evaluating ABS and TBS Performance in Urban Environments}
The primary objective of this section is to evaluate the performance of ABS and TBS, resulting in a better understanding of coverage and service quality for vehicles in urban environments. For this, we simulate the deployment of ABSs and TBSs using stochastic models and analyze various performance parameters, such as the Signal-to-Noise Ratio (SNR) experienced by vehicles moving along roads.

\subsection{Simulation Scenario}
\begin{figure}[!t]
\begin{center}
  \includegraphics[width= 1\linewidth]{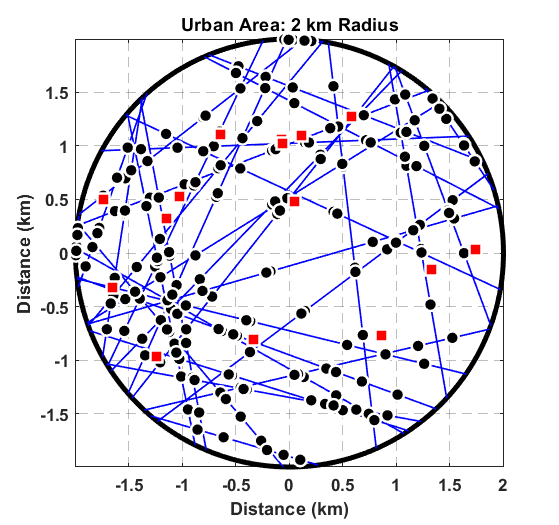}
  \caption{Simulation scenario where red square represents BS locations with intensity $\lambda_\chi$, blue lines are road with intensity $\lambda_{\text{road}}$, and black circles are vehicles on the road.}
  \label{layout}
  \end{center}
\end{figure}

A Poisson Point Process (PPP) is used to model wireless node positions in a defined area with intensity $\lambda$ \cite{ppp}. This intensity defines the average node points per $km^2$. Therefore, this paper places the ABSs and TBSs in a circular space with radius $r$, as shown in Fig. \ref{layout}. For a disk of radius $r$, the expected number of Base Stations (BSs) with intensity $\lambda_{\chi}$, where $\chi \in \{ \text{TBS}, \text{ABS} \}$, are:

\begin{equation}
\mathbb{E}[\chi] = \lambda_{\chi} \cdot \pi r^2
\end{equation}

The actual number of BSs $(n_{\chi})$, is generated from a Poisson distribution:
\begin{equation}
n_{\chi} \sim \text{Poisson}(\mathbb{E}[\chi])
\end{equation}
The BS locations are uniformly sampled within a disk using polar coordinates $(\rho, \theta)$, where $\theta \sim \mathcal{U}(0, 2\pi)$ and $\rho \sim \mathcal{U}(0, r)$. The Cartesian coordinates are given by:
\begin{equation}
x_{\chi} = \rho \cdot \cos(\theta),~~~ y_{\chi} = \rho \cdot \sin(\theta)
\end{equation}

The PPP generates random points in the defined area, making it less suitable for positioning vehicles because their location is restricted to roadways. Therefore, we use a Poisson Line Process (PLP) to model random roads in the same area with intensity $\lambda_{\text{road}}$ \cite{PLP}. The expected number of roads within the disk is given by:
\begin{equation}
\mathbb{E}[\text{road}] = 2 \pi r \cdot \lambda_{\text{road}}
\end{equation}
The number of roads $\mathcal{L}$ drawn using PLP are:
\begin{equation}
\mathcal{L} \sim \text{Poisson}(\mathbb{E}[\text{road}])
\end{equation}

Each road line is defined by two endpoints, computed based on a random sampling of angles and distances within the disk. The endpoints of each line are determined using:
\begin{equation}
x_1 = p \cdot \cos(\theta) + q \cdot \sin(\theta), ~~~ y_1 = p \cdot \sin(\theta) - q \cdot \cos(\theta)
\end{equation}
\begin{equation}
x_2 = p \cdot \cos(\theta) - q \cdot \sin(\theta), ~~~ y_2 = p \cdot \sin(\theta) + q \cdot \cos(\theta)
\end{equation}

where $p = r \cdot u_i, \quad u_i \sim \mathcal{U}(0, 1), \quad i = 1, 2, \dots, n_{\text{lines}}$ and $q = \sqrt{r^2 - p_i^2}, \quad i = 1, 2, \dots, n_{\text{lines}}$ are distances defining the line segment, and $\theta$ is the angle of the line.

This case study considers 200 cars randomly distributed over the lines/roads. Next, the 2D distance ($d_{\text{2D}}$) and 3D distance ($d_{\text{3D}}$) to all BSs is computed for each vehicle to determine the nearest base station serving $ith$ vehicles on $mth$ road ($v_{im}$), as highlighted in Algorithm \ref{algo1}. 

\begin{algorithm}[!t]

\caption{ Performance Analysis of ABS and TBS using Poisson Processes for ITS.}
\label{algo1}
\begin{algorithmic}[1]
\STATE \textbf{Input:} $\lambda_{\text{TBS}}, \lambda_{\text{ABS}}, \lambda_{\text{road}}, f, G_{tx}, G_{rx}, BW, k, T, N$ 

\STATE \textbf{Initialize:} Arrays for $PL$ and SNR for ABS and TBS.

\STATE \textbf{Generate Base Stations:}
\STATE Deploy $n_{\text{TBS}} \sim \text{PPP}(\lambda_{\text{TBS}})$, $n_{\text{ABS}} \sim \text{PPP}(\lambda_{\text{ABS}})$

\STATE \textbf{Generate Roads and Vehicles:}
\STATE Generate roads $\mathcal{L}_m \sim \text{PLP}(\lambda_{\text{road}})$, where $m$ denotes the index of the road
\STATE Place vehicles $\{v_{im}\}$ uniformly along each road $\mathcal{L}_m$

\STATE \textbf{For each vehicle $v_{im}$ on road $\mathcal{L}_m$:}
\STATE Restrict $v_{im}$ to its assigned road segment $\mathcal{L}_m$

\FOR{each BS type $j \in \{\text{TBS}, \text{ABS}\}$}
    \STATE Compute distances $d_{ijm}$ from $v_{im}$ to each BS in $n_{j}$
    \STATE Determine the closest BS for vehicle $v_{im}$:
    \[
    \text{BS}_{\text{serve},jim} = \arg\min_{b \in n_{j}} d_{ijm}(b)
    \]
    \STATE Calculate $\mathcal{P}_{\text{LoS}}$ and $PL$ for each  $(v_{im}, \text{BS}_{\text{serve},jim})$
    \STATE Compute and store SNR for each $v_{im}$f
\ENDFOR

\STATE \textbf{Repeat:} Perform for $N$ simulations for averaging

\STATE \textbf{Post-Process:} Performance parameters

\end{algorithmic}
\end{algorithm}

\subsection{Performance Metrics}
The Path Loss ($PL^\chi$) is computed based on the proportion of LoS and Non-LoS (NLoS) channels using 
\begin{equation}
\label{PL}
    PL^\chi = \mathcal{P}_{\text{LoS}}^\chi \cdot PL_{\text{LoS}}^\chi +  (1-\mathcal{P}_{\text{LoS}}^\chi) \cdot PL_{\text{NLoS}}^\chi,
\end{equation}
where $\mathcal{P}_{\text{LoS}}$ represents LoS probability, while $PL_{LoS}$ and $PL_{NLoS}$ are path losses for LoS and NLoS links, respectively. We consider an Urban Macro (UMa) environment defined by the 3rd Generation Partnership Project (3GPP) \cite{3gpp}. The $\mathcal{P}_{\text{LoS}}$ for TBS ($\mathcal{P}_{\text{LoS}}^T$) and ABS ($\mathcal{P}_{\text{LoS}}^A$) in UMa can be estimated using equations \eqref{plosT} and \eqref{plosA}.

\begin{equation}
\label{plosT}
    \mathcal{P}_{\text{LoS}}^T = 
\begin{cases}
1 , & d_{\text{2D}} \leq 18 \text{ m}, \\
\frac{18}{d_{\text{2D}}} + \exp\left(-\frac{d_{\text{2D}}}{63}\right)\left(1 - \frac{18}{d_{\text{2D}}}\right), & 18 \text{ m} < d_{\text{2D}},
\end{cases}
\end{equation}

\begin{equation}
\label{plosA}
    \mathcal{P}_{\text{LoS}}^A = 
\begin{cases}
1, & d_{\text{2D}} \leq d_1, \\
\frac{d_1}{d_{\text{2D}}} + \exp\left(-\frac{d_{\text{2D}}}{p_1}\right)\left(1 - \frac{d_1}{d_{\text{2D}}}\right), & d_{\text{2D}} > d_1,
\end{cases}
\end{equation}
where $p_1 = 4300\log_{10}(h_{ABS}) - 3800$ and $d_1 = max(460\log_{10}(h_{ABS}) - 700, 18)$, and $h_{ABS}$ is ABS height.  

The $PL_{LoS}^\chi$ and $PL_{NLoS}^\chi$ can be estimated using the following equations \cite{3gpp, 3gppGBS}:
\begin{equation}
    PL_{\text{LoS}}^\chi = 28 + 22 \log_{10}(d_{\text{3D}}) + 20 \log_{10}(f),
\end{equation}
\begin{equation}
    PL_{\text{NLoS}}^\chi = 
    \begin{cases}
    32.4 + 30 \log_{10} (d_{\text{3D}}) + 20 \log_{10}(f), \chi = TBS \\
    \begin{aligned}
    &-17.5 + (46 - 7\log_{10}(h_{\text{ABS}}))\log_{10}(d_{\text{3D}})  \\
    &\quad+ 20\log_{10}\left(\frac{40\pi f}{3}\right), \text{if } \chi = ABS
    \end{aligned} 
    \end{cases}
\end{equation}
where $f$ is the frequency in GHz. The Signal-to-Noise Ratio $SNR^\chi$ is computed using the formula:
\begin{equation}
\label{snr}
\text{SNR}^\chi = P_{\text{tx}}^\chi + G_{\text{tx}}^\chi + G_{\text{rx}}^\chi - PL^\chi - 10 \log_{10}(k T BW),
\end{equation}
where $P_{\text{tx}}$ is the transmission power in dBm, $G_{\text{tx}}$ and $G_{\text{rx}}$ are transmitter and receiver gains in dBi, $k$ is the Boltzmann constant ($1.38 \times 10^{-23} J/K$), $T$ is the temperature in Kelvin, and $BW$ is the bandwidth in Hz. 

Similarly, The Coverage Probability $\mathcal{P}_{Cov}^\chi(\gamma)$ and Outage Probability $\mathcal{P}_{Out}^\chi(\gamma)$ for TBS and ABS at a given SNR threshold $\gamma$ is calculated as follows:
\begin{equation}
\mathcal{P}_{Cov}^\chi(\gamma) = \mathbb{P}\left(\text{SNR}^\chi > \gamma\right) = \frac{1}{N} \sum_{i=1}^{N} \mathbb{I} \left( \text{SNR}_i^\chi > \gamma \right),
\end{equation}
\begin{equation}
\mathcal{P}_{Out}^\chi(\gamma) = \mathbb{P}\left(\text{SNR}^\chi < \gamma\right) =  \left( 1 - \mathcal{P}_{Cov}^\chi(\gamma)  \right),
\end{equation}
where $\mathbb{I}(\cdot)$ is the indicator function, and $N$ is the total number of $SNR$ samples. Finally, the Coverage Area $A_{cov}$ as a function of the $SNR$ threshold $\gamma$ is given by
\begin{equation}
A_{cov}^\chi(\gamma) = \pi \left(r \sqrt{\mathbb{P}(\text{SNR}^\chi > \gamma)}\right)^2.
\end{equation}

\subsection{Performance Evaluation}
This study assumes 5G BSs with perfect beamforming, hence completely mitigating the impact of interference. The simulation parameters include $G_{\text{tx}}$ = 10~dBi, $G_{\text{rx}}$ = 5~dBi, $P_{\text{tx}}= 28$~dBm, $f$ = 28~GHz, and $BW$ = 100~MHz. It is assumed that the closest BS will serve the cars (200 in total), and each simulation is repeated 500 times for averaging.

\begin{figure}[!t]
\begin{center}
  \includegraphics[width=1\columnwidth]{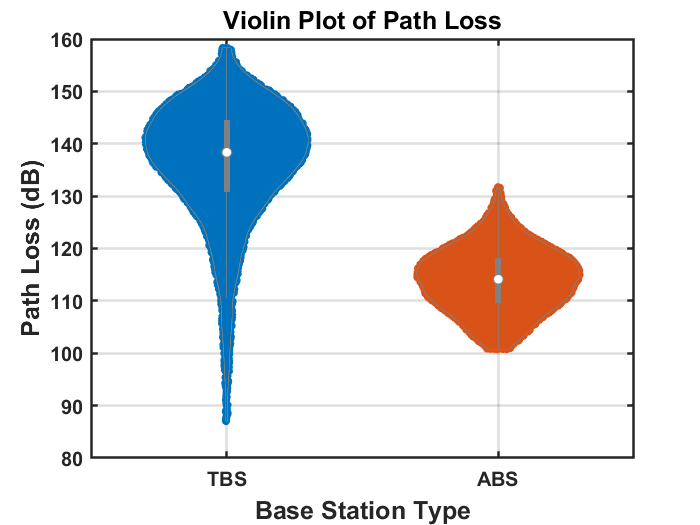}
  \caption{$PL$ distribution for ABS and TBS.}
  \label{PLv}
  \end{center}
\end{figure}

Fig. \ref{PLv} compares the $PL$ distribution for ABS (at 100~m) and TBS in an urban area with $r$ = 2~km and $\lambda_\chi$ = 2. The violin shape of the figure represents the $PL$ probability density, with the wider parts indicating higher probability. From the figure, ABS demonstrates a lower median $PL$ with a few outliers, demonstrating its ability to provide stable connectivity to vehicles. In contrast, TBS exhibits a broader distribution of $PL$ with a higher median $PL$ than ABS. However, it can achieve peak $PL$ values due to lower 3D distance than ABS. 

\begin{figure}[!t]
  \centering
    \begin{subfigure}[b]{0.9\linewidth}
    \includegraphics[width=\linewidth]{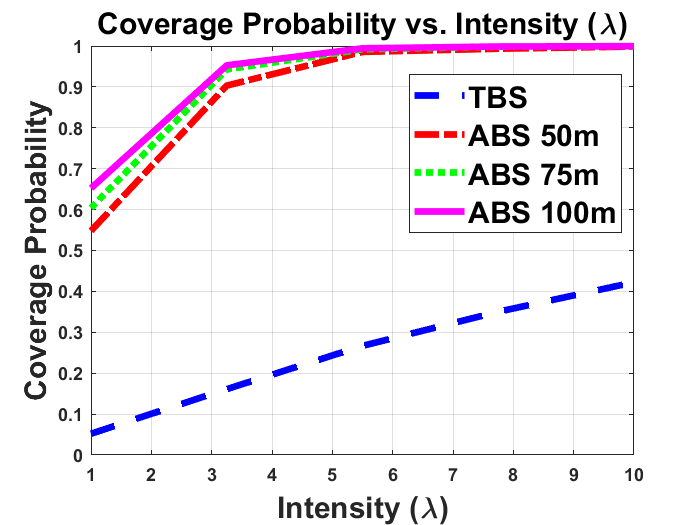}
    \caption{$\gamma$ = 15~dB }
    \label{Cvsh15}
  \end{subfigure}
  \centering
  \begin{subfigure}[b]{0.9\linewidth}
    \includegraphics[width=\linewidth]{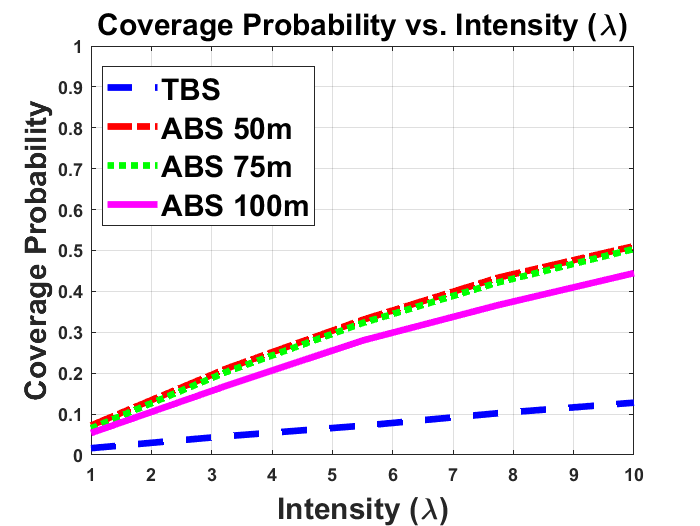}
    \caption{$\gamma$ = 30~dB}
    \label{Cvsh30}
  \end{subfigure}
  \centering
  \caption{$\mathcal{P}_{Cov}$ for ABS/TBS against varying $\lambda_\chi$ and $h_{\text{ABS}}$.}
  \label{Cvsh}
\end{figure}

Fig. \ref{Cvsh} compares $\mathcal{P}_{Cov}$ for ABS and TBS against varying $\lambda_\chi$ and $h_{\textbf{ABS}}$. In Fig. \ref{Cvsh15}, the SNR threshold $\gamma$ is set to 15 dB, a minimum requirement to establish a connection. The results indicate that ABSs offer a significantly higher $\mathcal{P}_{Cov}$ than TBSs, especially at higher altitudes due to the better availability of LoS. The figure also illustrates that increasing the ABS altitude enhances $\mathcal{P}_{Cov}$ when the BS density is low. However, as $\lambda$ increases, the influence of altitude starts reducing because the environment becomes saturated with sufficient BSs to handle the traffic.
In contrast, higher ABS altitudes incur higher $PL$ and reduced SNR due to increased $d_{\text{3D}}$. Therefore, elevated altitudes may not produce optimal results for scenarios with higher $\gamma$ requirements, as demonstrated in Figure \ref{Cvsh30}.

\begin{figure}[!t]
   \begin{minipage}{0.24\textwidth}
     \centering
     {\includegraphics[width=1\linewidth]{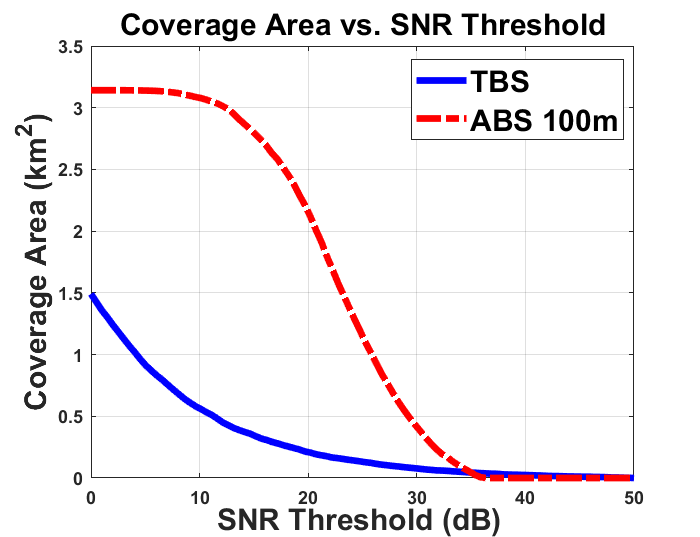}}
     \caption{$A_{cov}$ vs. varying $\gamma$.}
     \label{CovA}
   \end{minipage}\hfill
   \begin{minipage}{0.24\textwidth}
     \centering
     {\includegraphics[width=1\linewidth]{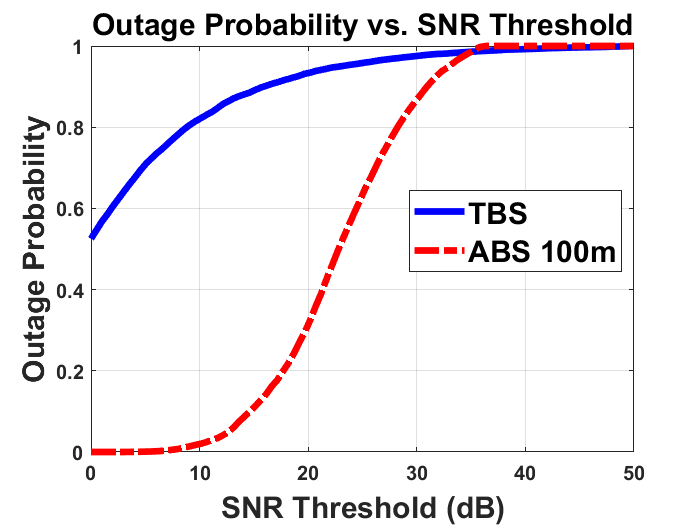}}
     \caption{$\mathcal{P}_{Out}$ vs. varying $\gamma$.}
     \label{outage}
   \end{minipage}
\end{figure}

\begin{figure}[!t]
\begin{center}
  \includegraphics[width=.9\columnwidth]{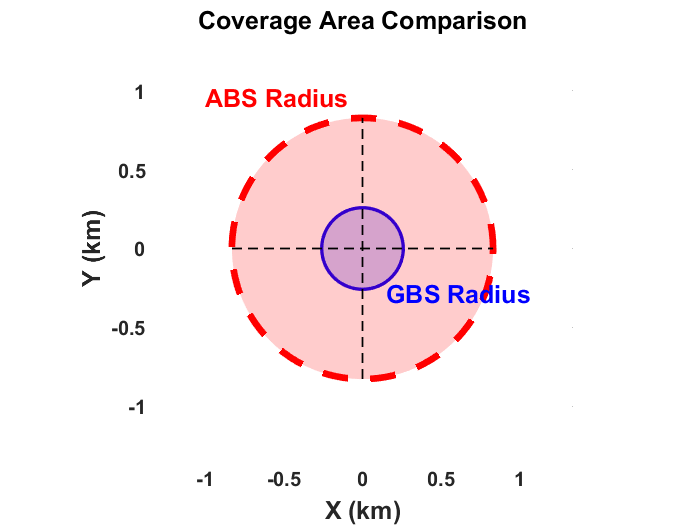}
  \caption{Coverage area visualization for $\lambda_\chi$ = 2.}
  \label{CovC}
  \end{center}
\end{figure}

Fig. \ref{CovA} demonstrates $A_{cov}$ against varying $\gamma$ for $\lambda$ = 2. As previously noted, higher altitudes for ABS significantly enhance coverage probability, leading to greater area coverage. For example, at $\gamma$ = 0, ABS at 100~m can cover $\approx$ 3.1 km$^2$ area compared to TBS covering only 1.5 km$^2$. 
However, as $\gamma$ increases, $A_{cov}$ for ABS begins to reduce from 3.1 km$^2$ to less than a km$^2$ due to the increased $PL$ associated with higher altitudes, which become more prominent at higher SNR thresholds. As a result, the effectiveness of ABSs in covering large areas decreases with increased SNR requirements. This highlights a trade-off between altitude and coverage performance that must be carefully managed depending on the target SNR threshold. Fig. \ref{CovC} visualizes the $A_{cov}$, and Fig.~\ref{outage} displays the $\mathcal{P}_{Out}$ associated with ABS and TBS at for $\lambda_\chi$ = 2 and $r$ = 2~km.

\begin{figure*}[!t]
\begin{center}
  \includegraphics[width=.8\linewidth]{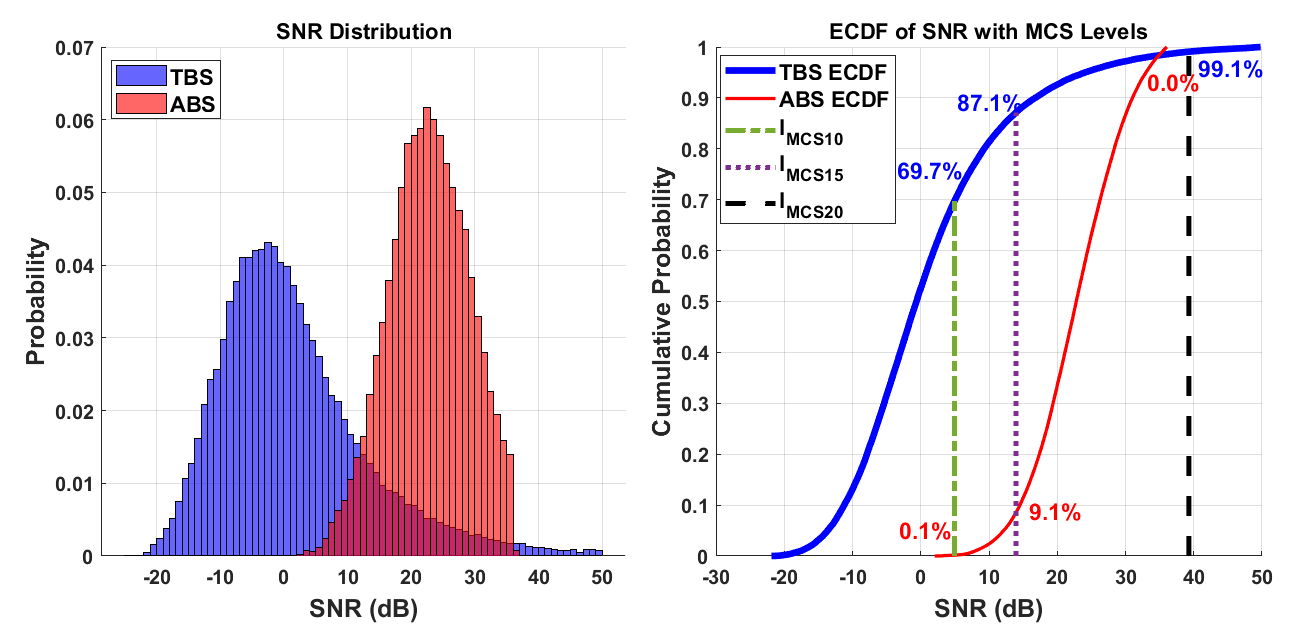}
  \caption{SNR distributions and ECDFs for TBS and ABS systems, with MCS thresholds indicated.}
  \label{CDF}
  \end{center}
\end{figure*}

Fig. \ref{CDF} presents the SNR distributions and Empirical Cumulative Distribution Functions (ECDFs) for ABS and TBS. The histogram on the left shows ABS achieves overall better SNR distribution, mainly ranging between 10~dB and 30~dB, whereas TBS exhibits lower SNR values, where most values reside between 0~dB and 10~dB. We observe a truncation around 37~dB for ABS, highlighting the maximum possible SNR achievable with the ABS. In contrast, a small proportion of GBS provides better performance when the link is LoS, and the 3D distance is lower. The ECDFs on the right further highlight ABS superiority, where it consistently surpasses the SNR thresholds required for higher Modulation and Coding Scheme (MCS) levels in 5G. For comparison, we plot the SNR thresholds (4.94~dB, 13.95~dB, 39.28~dB) for three different MCS indexes ($I_{MCS10}$, $I_{MCS15}$, $I_{MCS20}$) for modulation orders 4,6 and 8, respectively \cite{3gpp5G}. For $I_{MCS10}$, only 0.1\% SNR values fall below the threshold for ABS compared to 69.7\% values of TBS. Hence, ABS has a 99.99\% probability of supporting $I_{MCS10}$ requirements than TBS with a 30.30\% probability. Similarly, ABS performs better for $I_{MCS15}$ but lags behind TBS for $I_{MCS20}$ where SNR threshold values are too high, which are difficult to achieve considering $PL$ due to higher 3D distance. However, this limitation can be mitigated by adjusting $h_{ABS}$, as illustrated in Figure \ref{Cvsh30}. In conclusion, the results highlight ABS's superior signal quality, making it more suitable for high-throughput ITS applications requiring advanced MCS levels.

Despite the advantages of ABSs, there are several challenges that limit their large-scale deployment. The first is the short battery life, which can be resolved using enhanced lithium-ion batteries, optimal UAV design, or energy harvesting techniques \cite{mohsan2022towards}. The second critical challenge is security and privacy, requiring advanced, lightweight encryption algorithms. Lastly, ensuring safety standards, airspace management, and other regulatory constraints, such as flight height and distance from people/sensitive areas, are bug concerns. Hence, there is an urgent need for collaboration between aviation authorities and governments to establish guidelines to ensure safe and efficient UAV integration in ITS.   

\section{Conclusion}
This paper envisions a 3D ITS by integrating UAV-based AAM and connectivity to address the limitations of existing 2D ITS, particularly in urban environments. First, it discusses the potential use cases of 3D ITS that can improve transportation and road user experiences. Next, it evaluates the performance of ABS and TBS in urban environments. The results demonstrate that ABSs offer better coverage and more reliable connectivity than TBS due to the ability to provide stable LoS connections over larger areas. This work sets the foundation for the next generation of ITS, which will support more dynamic, flexible, and resilient transportation networks. In the future, we will explore optimizing ABS altitude, locations, and beamforming strategies to enhance coverage and reduce interference. 

\section*{Acknowledgment}
This research is supported by the Research Foundation Flanders (FWO), project no. G098020N, and iSEE-6G project under the Horizon Europe Research and Innovation program with Grant Agreement No. 101139291.

\balance
\bibliographystyle{IEEEtran}
\bibliography{references}

\end{document}